\documentclass[aps,prl,twocolumn,showpacs,preprintnumbers,amsmath,amssymb]{revtex4}

\usepackage{graphicx}

\begin{document}

\title{Impact of interactions on human dynamics}

\author{J. G. Oliveira$^{1}$ and A. Vazquez$^{2}$}

\affiliation{$^1$ Departamento de F\'{\i}sica, Universidade de Aveiro, 3810-193, Aveiro, Portugal}
\affiliation{$^2$ The Simons Center for System Biology,
The Institute for Advanced Study, Einstein Dr, Princeton, NJ 08540, USA}

\date{\today}

\begin{abstract}

Queueing theory has been recently proposed as a framework to model the 
heavy tailed statistics of human activity patterns. The main predictions 
are the existence of a power-law distribution for the interevent time of 
human actions and two decay exponents 
$\alpha=1$ and $\alpha=3/2$. Current models lack, however, a key aspect of 
human dynamics, i.e. several tasks require, or are determined by, 
interactions between individuals. Here we introduce a minimal queueing 
model of human dynamics that already takes into account human-human 
interactions. To achieve large scale simulations we obtain a 
coarse-grained version of the model, allowing us to reach large interevent times and reliable scaling exponents estimations. Using this 
we show that the interevent distribution of interacting tasks exhibit the 
scaling exponents $\alpha=2$, 3/2 and a series of numerable values between 
3/2 and 1. This work demonstrates that, within the context of queueing 
models of human dynamics, interactions change the exponent of the power-law distributed interevent times. Beyond the study of human dynamics, these results are relevant to systems 
where the event of interest consists of the simultaneous occurrence of 
two (or more) events.

\end{abstract}

\pacs{89.75.Da,02.50.Le,89.65.Ef,89.75.Hc}

\maketitle


Understanding the timing of human activities is extremely important to
model human related activities such as communication systems
\cite{reynolds03} and the spreading of computer viruses \cite{vazquez07}.
In the recent years we have experienced an increased research activity in
this area motivated by the increased availability of empirical data. We
now count with measurements of human activities covering several
individuals and several events per individual
\cite{ebel02,eckmann04,barabasi05,oliveira05,dezso06}. Thanks to this data we are in
a position to investigate the laws and patterns of human dynamics using a
scientific approach.

Barab\'asi has taken an important step in this direction reconsidering
queueing theory \cite{cohen69,gross98} as framework to model human dynamics \cite{barabasi05}.
Within this framework, the {\em to do list} of an individual is modeled as
a finite length queue with a task selection protocol, such as highest
priority first. The main predictions are the existence of a power law
distribution of interevent times $P_\tau\sim\tau^{-\alpha}$ and two
universality classes characterized by exponents $\alpha=1$
\cite{barabasi05,vazquez05PRL,vazquez06PRE} and $\alpha=3/2$
\cite{oliveira05,vazquez06PRE}. These universality classes have been
corroborated by empirical data for email \cite{barabasi05,vazquez06PRE}
and regular mail communications \cite{oliveira05,vazquez06PRE},
respectively, motivating further theoretical research
\cite{grinstein06PRL,gabrielli07PRL,Blanchard:2007fk}.

The models proposed so far have been limited, however, to single 
individual dynamics. In practice people are connected in social networks 
and several of their activities are not performed independently. This 
reality forces us to model human dynamics in the presence of interactions 
between individuals. Our past experience with phase transitions has shown 
us that interactions and their nature are a key factor determining the 
universality classes and their corresponding scaling exponents 
\cite{stanley71}. Furthermore, beyond the study of human dynamics, there 
are several systems where the event of interest consists of the 
simultaneous occurrence of two (or more) events. For example, collective 
phenomena in disordered media, such as the interaction of two (or more) particles in cluster formation.

\begin{figure}[t]

\includegraphics[width=2.1in]{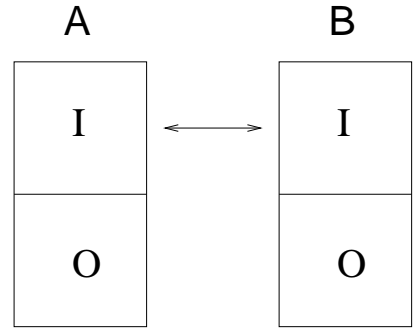}

\caption{System of two agents with a common interacting task I and an
aggregate task O representing a set of individual tasks.}

\label{f:2queues}
\end{figure}

To investigate the impact of human-human interactions on the timing of
their activities we consider a minimal model consisting of two agents, A
and B (Fig. \ref{f:2queues}). Each agent is modeled by a priority list
containing two tasks, interacting task (I) and aggregate non-interacting
task (O). The interacting task models a common activity such as meeting
each other, requiring the simultaneous execution of that task by both
agents. On the other hand, the non-interacting task represents an
aggregate meta-activity accounting for all other tasks the agents execute,
which do not require an interaction between them. To each task we assign
random priorities $x_{ij}$ ($i=I,O$; $j=A,B$) extracted from a probability
density function (pdf) $f_{ij}(x)$ (see Fig. \ref{f:2queues}).

The rules governing the dynamics are as follows. {\it Initial condition:}
We start with a random initial condition, assigning a priority to the I
and O tasks from their corresponding pdf. {\it Updating step:} At each
time step, both agents select the task with higher priority in their list.
If (i) both agents select the interacting task then it is executed, (ii)
otherwise each agent executes the O task, representing the execution of
any of their non-interacting tasks.

\begin{figure}[t]
\includegraphics[width=3.1in]{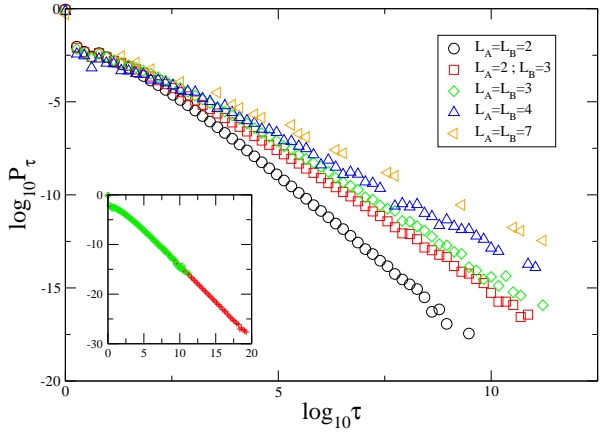}

\caption{Probability distribution of the interevent time $\tau$ of the
interacting task I, as obtained from the direct numerical simulations of
the model. Each dataset was obtained after $10^{11}$ model time steps,
corresponding with total number of I plus O task executions. Note that as
$L_A$ and/or $L_B$ increases it becomes computationally harder to have a
good estimate of $P_\tau$ because the execution of the I task becomes less
frequent. The inset shows the distribution for $L=3$ as obtained from the 
original model with $10^{12}$ steps (green diamonds), and the 
coarse-grained model with $N=10^9$ (red plus), derived to obtain more reliable estimation of the exponents.}

\label{f:original}
\end{figure}

Our aim is to determine the impact of the interaction between the agents
and the shape of $f_{ij}(x)$ on the scaling exponent $\alpha$ of the
interevent time distribution of the interacting task I. For simplicity, we
focus on the following priority distribution. Consider the case where each
agent has $L_j$ ($j=A,B$) tasks, one I task and $L_j-1$ non-interacting
tasks, their priorities following a uniform distribution in the interval
$[0,1]$. The pdf of the highest priority among $L_j-1$ tasks is in this
case given by $(L_j-1)x^{L_j-2}$, resulting in

\begin{equation}\label{e:fij_uniform}
f_{ij}(x) = \left\{
\begin{array}{ll}
1\ ,& i=I\\
(L_j-1)x^{L_j-2}\ , & i=O\ .
\end{array}
\right.
\end{equation}

\noindent This example shows that the priorities pdf of task I and O are
in general different. All the results shown below were obtained using the
pdf in Eq. (\ref{e:fij_uniform}).

\begin{figure}[t]
\includegraphics[width=3.1in]{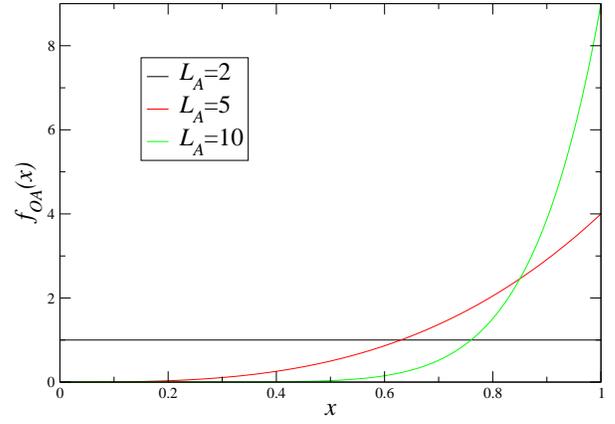}

\caption{Probability density function of the non-interacting aggregate
task priority of user A, as obtained from Eq. (\ref{e:fij_uniform}). With
increasing the queue length $L_A$, $f_{OA}(x)$ concentrates more and more
in the vicinity of $x=1^{-}$.}

\label{f:Opdf}
\end{figure}

To investigate the interevent time distribution we perform extensive
numerical simulations. Figure \ref{f:original} shows the interevent time
distribution as obtained from direct simulations of the model introduced
above. It becomes clear that for large $L_A$ and/or $L_B$ we do not obtain
a good statistics, even after waiting for $10^{11}$ updating steps. This
observation is a consequence of the behavior of $f_{Oj}(x)$ when $L_A$
and/or $L_B$ are large (Fig. \ref{f:Opdf}). Focusing on agent A, as $L_A$
increases $f_{OA}(x)$ gets more concentrated around priority one,
while the priority of the I task remains uniformly spread between zero and
one. This fact results in increasingly large interevent times between the
execution of the I task.

To speed-off the numerical simulations we derive a coarse-grained version
of the model, allowing us to analyze the scaling behavior of the
interevent time distribution over several orders of magnitude (inset of Fig. \ref{f:original}). We start by
noticing that, given $(x_{IA},x_{IB})$, the joint pdf of $(x_{OA},x_{OB})$
factorizes and the probability $q(x_{IA},x_{IB})$ that both agents execute
I right after O is given by

\begin{equation}\label{e:q}
q(x_{IA},x_{IB}) = \int_0^{x_{IA}} dx f_{OA}(x)
\int_0^{x_{IB}} dx f_{OB}(x) \ .
\end{equation}

\noindent This factorization is possible because the execution of the I
task requires its priority to be the largest for both agents.
In turn, with probability $1-q(x_{IA},x_{IB})$ both agents
continue to execute O. Thus, the probability distribution
$Q_\tau(x_{IA},x_{IB})$ that I waits $\tau>1$ steps before being executed
follows the geometric distribution

\begin{equation}\label{e:Qtau}
Q_{\tau}(x_{IA},x_{IB}) = q(x_{IA},x_{IB})
[1-q(x_{IA},x_{IB})]^{\tau-2} \ .
\end{equation}

\noindent Once the I task is executed it can be executed again resulting
in interevent times of one step ($\tau=1$). The overall interevent time
distribution of the I task is given by

\begin{equation}\label{e:Ptau}
P_\tau = \left\{
\begin{array}{ll}
P_1\ ,& \tau=1\\
(1-P_1)\langle Q_{\tau}(x_{IA},x_{IB})\rangle\ , & \tau>1
\end{array}
\right.
\end{equation}

\noindent where

\begin{equation}\label{e:P1}
P_1 = \frac{S_1}{S_1+1}\ ,
\end{equation}

\noindent $S_1$ is the expected number of consecutive executions of the I
task and $\langle\cdots\rangle$ denotes the expectation over different
realizations of $(x_{IA},x_{IB})$, just at the step of switching from task
I to O. Finally, at the step of switching from O to I, the O task priority
of both agents must fall below that of the I task. Therefore, the pdf of
$x_{Oj}$ ($j=A,B$) just after the switch from O to I is given by

\begin{equation}\label{e:pdf*}
f_{Oj}^*(x|x_{Ij})=\frac{f_{Oj}(x)}{\int_0^{x_{Ij}} f_{Oj}(x') dx'} \ ,
\end{equation}

\noindent where $0\leq x<x_{Ij}$. This later result together with Eq.
(\ref{e:Qtau}) allow us to condense all steps with consecutive
executions of the O task into a single coarse-grained step. More
important, this mapping is exact.

Putting all together the coarse-grained model runs as follows. {\it
Initial condition:} We start with random initial priorities extracted from
the pdfs $f_{ij}(x)$. {\it Updating step:} At each step, (i) if for both
agents the I task priority is larger than that for the O task we run the
model as defined above, both agents executing the I task and updating
their I task priorities using the pdfs $f_{Ij}$ ($j=A,B$). (ii) Otherwise,
we generate a random interevent time $\tau$ from the probability
distribution (\ref{e:Qtau}) and a new O task priority for each agent using
the pdf $f^*_{Oj}(x|x_{Ij})$ (\ref{e:pdf*}). This second step avoids going
over successive executions of the O task which, for a large number of
non-interacting tasks, significantly slow down the simulations.

The second step of the coarse grained model requires us to extract a
random number from a geometric distribution. This can be achieved very
efficiently exploiting the fact that the integer part of a real random
variable with an exponential distribution follows a geometric
distribution. Using this fact, when $\tau>1$, we extract $\tau$ exactly from
the distribution in Eq. (\ref{e:Qtau}), which differs from the
corresponding branch of Eq. (\ref{e:Ptau}). Normalization by the total
number of task I executions, including those with $\tau=1$, provides
$\tau>1$ distributed according to Eq. (\ref{e:Ptau}).

\begin{figure}[t]
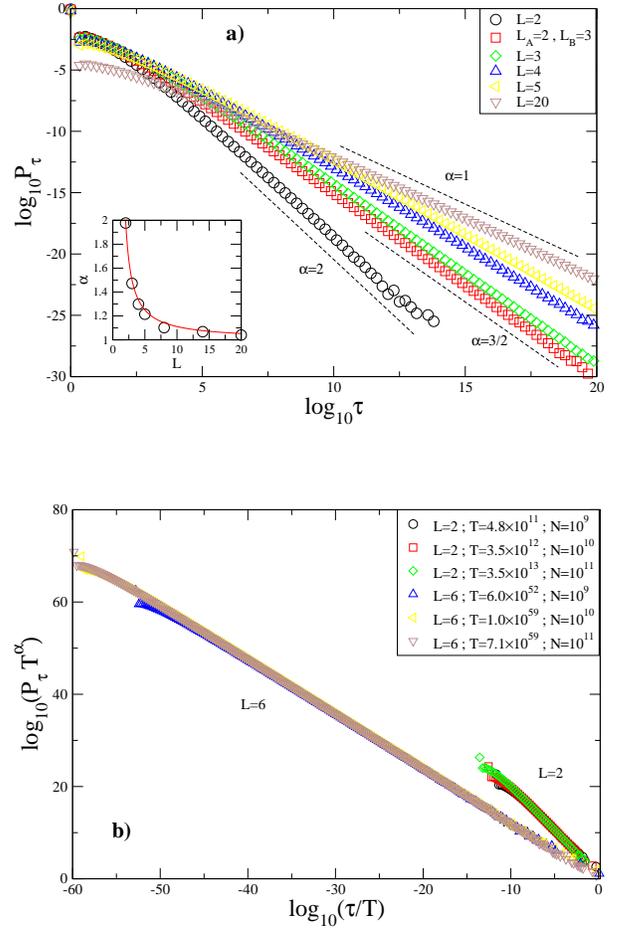


\includegraphics[width=3.1in]{Ptau_LaLb.eps}
\vspace{1cm}
\includegraphics[width=3.1in]{scaling_L2_L6.eps}

\caption{a) Probability distribution of the I task interevent time for
several values of the number of tasks on each queue ($L_A$, $L_B$), as
obtained from simulations of the coarse-grained model. When $L_A=L_B$ we
denote this number by $L$. The inset shows the exponent $\alpha$ as measured from the power
law tails (black circles) and the guess function $\alpha=1+1/\max(L_j-1)$ (red curve) in good agreement; to avoid confusion we only plot the case when $L_A=L_B=L$, but we checked for the general case as well. b) Scaling plot of the I task interevent
time distribution. Note that, for a given $\alpha$, the symbols
corresponding to different time windows $T$ collapse into a single plot.}

\label{f:c-g}
\end{figure}

The I task interevent time distribution obtained from simulations of the
coarse-grained model is plotted in Fig. \ref{f:c-g}a. When $L_A=L_B=L=2$
it follows a power-law tail with exponent $\alpha=2$. As $L$ increases $\alpha$
approaches one. A guess to this dependence, in good agreement with the measured values, is given by $\alpha=1+1/\max(L_j-1)$ (inset of Fig. \ref{f:c-g}a). The
numerical results indicate that there are several numerable universality
classes parameterized by $L_A$ and $L_B$. Notice that the second largest
value of $\alpha$ (obtained when $L_A=2$ and $L_B=3$, or vice-versa) is
close to $3/2$ and, therefore, our results do not show universality
classes with exponent $\alpha$ between 3/2 and 2 (unless we assume real
valued queue lengths).

The power laws in Fig. \ref{f:c-g}a exhibit a cutoff at a certain value of
$\tau$. To investigate if this is a natural cutoff or just a finite size
effect, we investigate the shape of the interevent time distribution as a
function of the observation time window $T$. The later is
defined as the total number
of steps considering both the I and O task and satisfy

\begin{equation}\label{e:T}
T=\sum_{i=1}^N\tau_i\ ,
\end{equation}

\noindent where $N$ is the number of executions of the I task within the
time window $T$ and $\tau_i$ ($i=1,\ldots,N$) is the sequence of
interevent times between executions of the I task. We assume that the
cutoff is determined by the finite time window and that the interevent
time distribution follows the scaling form

\begin{equation}\label{e:Ptau_scaling}
P(\tau)=A\tau^{-\alpha}g\left(\frac{\tau}{T^z}\right)
\end{equation}

\noindent where A is a constant, $z>0$ is a scaling exponent and $g(x)$ is
a scaling function with the asymptotic behaviors $g(x)\approx1$ when
$x\ll1$ and $g(x)\ll1$ when $x\gg1$. Under this assumption
$P(\tau)\sim\tau^{-\alpha}$ when $T\rightarrow\infty$, with
$1<\alpha\leq2$. Given this power law tail and exponent, the number of
interevent times $N$ necessary to cover the window $T$ is of the order of
magnitude of $T^{\alpha-1}$ \cite{feller71}. In turn, the mean intervent
time is of the order of

\begin{equation}\label{e:mean_tau}
\langle \tau\rangle = \frac{1}{N}\sum_{i=1}^N\tau_i \sim T^{2-\alpha}\ .
\end{equation}

\noindent From Eqs. (\ref{e:Ptau_scaling}) and (\ref{e:mean_tau}) it
follows that $z=1$.

To check our scaling assumption we plot $P_\tau T^{\alpha}$ as a function
of $\tau/T$ (Fig. \ref{f:c-g}b). The symbols corresponding to different
time windows $T$ clearly overlap into a single curve, demonstrating that
the scaling assumption in Eq. (\ref{e:Ptau_scaling}) is correct with
$z=1$. Thus, in the $T\rightarrow\infty$ limit the I task intervent time
distribution exhibits a true power law tail $P_\tau\sim\tau^{-\alpha}$.

Within the context of queuing models of human dynamics, only two 
universality classes were previously identified, corresponding to the 
single queue models of Cobham \cite{oliveira05,cobham54} ($\alpha=3/2$) 
and Barab\'asi \cite{barabasi05} ($\alpha=1$). The analysis of the two 
interacting agents model reveals that that the interaction between agents 
results in a richer set of exponents. Our numerical results provide 
evidence of a new universality class with exponent $\alpha=2$ and 
exponents between 3/2 and 1. It is worth noticing that the exponents 2 and 
1 may also result from a Poisson proccesses with a time dependent rate 
\cite{hidalgo06,vazquezPA07}.

Because the exponent $\alpha$ depends on the systems details, here 
represented by the agent's queue lengths $L_A$ and $L_B$, we conclude that 
the model with two interacting agents exhibits non-universal behavior. 
Interestingly, the exponent $\alpha=1$ is asymptotically reached when the 
number of tasks of one or both agents becomes large. As humans get engaged 
in several tasks this later asymptotic behavior may explain the ubiquitous 
observation of the exponent $\alpha=1$ \cite{vazquez06PRE}.

We use the number of non-interacting tasks as a mean to modulate the
distribution of the non-interacting aggregate task priority. Yet, it
is the distribution shape the primary factor determining the scaling
exponent $\alpha$. The effect of increasing $L_A$ and/or $L_B$ is a
concentration of the non-interacting aggregate task priority around
priority one, resulting in values of $\alpha$ that approaches one. This
means that the limit $\alpha=1$ is achieved for low priority interacting
tasks that remain most of the time in the queue without being executed, at
expenses of the execution of tasks which in general have a higher
priorities.

Considering the interaction between agents we also solve one of the
longstanding problems of the original single queue Barab\'asi model, related to the
stationarity of the interevent time distribution
\cite{vazquez05PRL,gabrielli07PRL}. In the Barab\'asi single queue model
the task with highest priority is executed with a probability $p$,
otherwise a task is selected at random for execution. When $p$ is close to
one the interevent time distribution exhibits a peak at one step and
$P_1\rightarrow1$ when $p\rightarrow 1$. When $p=1$ the distribution is
non-stationary and $P_1\rightarrow1$ when time $t\rightarrow\infty$. In
contrast, in the model considered here there is no need to introduce the
random selection rule and the corresponding model parameter $p$. The
interacting task interevent time distribution is stationary even when the
- highest priority first - selection rule is applied. In turn, the
exponent $\alpha$ is not exactly one, but reaches one asymptotically with
increasing the number of tasks. Finally, the interevent time distribution
of the Barab\'asi model exhibits a natural cutoff determined by the
parameter $p$, while for the model introduced here it is a true power law up to finite size effects.
However, it is worth noticing that in a recent work~\cite{goncalves08} it has been found that the original Barab\'asi's model with variable task execution rate can generate interevent time distributions with exponent $\alpha=1.25$. In principle, in our model also different choice of model parameters could result in other exponents in the range between 1 and 3/2.

This work represents the first step in understanding how interactions
among agents affect their activity pattern. Based on recent works using queueing theory we describe the model in the context of human dynamics. It can be generalized to
consider a larger number of agents connected by a specific social network. Also, the model can potencially be used more generally to study the time statistics of events requiring synchronization between two physical systems.

We thank J.F.F. Mendes for useful comments and suggestions.


\begin{thebibliography}{12}
\expandafter\ifx\csname natexlab\endcsname\relax\def\natexlab#1{#1}\fi
\expandafter\ifx\csname bibnamefont\endcsname\relax
  \def\bibnamefont#1{#1}\fi
\expandafter\ifx\csname bibfnamefont\endcsname\relax
  \def\bibfnamefont#1{#1}\fi
\expandafter\ifx\csname citenamefont\endcsname\relax
  \def\citenamefont#1{#1}\fi
\expandafter\ifx\csname url\endcsname\relax
  \def\url#1{\texttt{#1}}\fi
\expandafter\ifx\csname urlprefix\endcsname\relax\def\urlprefix{URL }\fi
\providecommand{\bibinfo}[2]{#2}
\providecommand{\eprint}[2][]{\url{#2}}

\bibitem{reynolds03} P. Reynolds, Call Center Staffing (The Call Center
School Press, Lebanon, Tennessee, 2003).

\bibitem{vazquez07} A. Vazquez, B. R\'acz, A. Luk\'acs and A.-L.
Barab\'asi, Phys. Rev. Lett. {\bf 98}, 158702 (2007).

\bibitem{ebel02} H. Ebel, L.-I. Mielsch, and S. Bornholdt, Phys. Rev. E
{\bf 66}, R35103 (2002).

\bibitem{eckmann04} J.-P. Eckmann, E. Moses, and D. Sergi, Proc. Natl.
Acad. Sci. U.S.A. {\bf 101}, 14333 (2004).

\bibitem[{\citenamefont{Barab\'asi}(2005)}]{barabasi05}
\bibinfo{author}{\bibfnamefont{A.-L.} \bibnamefont{Barab\'asi}},
  \bibinfo{journal}{Nature (London)} \textbf{\bibinfo{volume}{435}},
  \bibinfo{pages}{207} (\bibinfo{year}{2005}).

\bibitem[{\citenamefont{Oliveira and Barab\'asi}(2005)}]{oliveira05}
\bibinfo{author}{\bibfnamefont{J.~G.} \bibnamefont{Oliveira}} \bibnamefont{and}
  \bibinfo{author}{\bibfnamefont{A.-L.} \bibnamefont{Barab\'asi}},
  \bibinfo{journal}{Nature (London)} \textbf{\bibinfo{volume}{437}},
  \bibinfo{pages}{1251} (\bibinfo{year}{2005}).

\bibitem{dezso06} Z. Dezso, E. Almaas, A. Luk\'acs, B. R\'acz, I.
Szakad\'at and A.-L. Barab\'asi, Phys. Rev. E {\bf 73}, 066132 (2006).

\bibitem[{\citenamefont{Cohen}(1969)}]{cohen69}
\bibinfo{author}{\bibfnamefont{J.~W.} \bibnamefont{Cohen}},
  \emph{\bibinfo{title}{The Single Server Queue}}
  (\bibinfo{publisher}{North-Holland, Amsterdam}, \bibinfo{year}{1969}).

\bibitem[{\citenamefont{Gross and Harris}(1998)}]{gross98}
\bibinfo{author}{\bibfnamefont{D.}~\bibnamefont{Gross}} \bibnamefont{and}
  \bibinfo{author}{\bibfnamefont{C.~M.} \bibnamefont{Harris}},
  \emph{\bibinfo{title}{Fundamentals of Queueing Theory}}
  (\bibinfo{publisher}{John Wiley \& Sons, New York}, \bibinfo{year}{1998}).

\bibitem[{\citenamefont{Vazquez}(2005)}]{vazquez05PRL}
\bibinfo{author}{\bibfnamefont{A.}~\bibnamefont{Vazquez}},
  \bibinfo{journal}{Phys. Rev. Lett.} \textbf{\bibinfo{volume}{95}},
  \bibinfo{pages}{248701} (\bibinfo{year}{2005}).

\bibitem[{\citenamefont{Vazquez et~al.}(2006)\citenamefont{Vazquez,
  Oliveira, Dezs\H{o}, Goh, Kondor, and Barab\'{a}si}}]{vazquez06PRE}
\bibinfo{author}{\bibfnamefont{A.}~\bibnamefont{Vazquez}},
  \bibinfo{author}{\bibfnamefont{J.~G.} \bibnamefont{Oliveira}},
  \bibinfo{author}{\bibfnamefont{Z.}~\bibnamefont{Dezs\H{o}}},
  \bibinfo{author}{\bibfnamefont{K.-I.} \bibnamefont{Goh}},
  \bibinfo{author}{\bibfnamefont{I.}~\bibnamefont{Kondor}}, \bibnamefont{and}
  \bibinfo{author}{\bibfnamefont{A.-L.} \bibnamefont{Barab\'{a}si}},
  \bibinfo{journal}{Phys. Rev. E} \textbf{\bibinfo{volume}{73}},
  \bibinfo{pages}{036127} (\bibinfo{year}{2006}).

\bibitem[{\citenamefont{Grinstein and Linsker}(2006)}]{grinstein06PRL}
\bibinfo{author}{\bibfnamefont{G.}~\bibnamefont{Grinstein}} \bibnamefont{and}
  \bibinfo{author}{\bibfnamefont{R.}~\bibnamefont{Linsker}},
  \bibinfo{journal}{Phys. Rev. Lett.} \textbf{\bibinfo{volume}{97}},
  \bibinfo{pages}{130201} (\bibinfo{year}{2006}).


\bibitem[{\citenamefont{Gabrielli and Caldarelli}(2007)}]{gabrielli07PRL}
\bibinfo{author}{\bibfnamefont{A.}~\bibnamefont{Gabrielli}} \bibnamefont{and}
  \bibinfo{author}{\bibfnamefont{G.}~\bibnamefont{Caldarelli}},
  \bibinfo{journal}{Phys. Rev. Lett.} \textbf{\bibinfo{volume}{98}},
  \bibinfo{pages}{208701} (\bibinfo{year}{2007}).

\bibitem[{\citenamefont{Blanchard and Hongler}(2007)}]{Blanchard:2007fk}
\bibinfo{author}{\bibfnamefont{P.}~\bibnamefont{Blanchard}} \bibnamefont{and}
  \bibinfo{author}{\bibfnamefont{M.-O.} \bibnamefont{Hongler}},
  \bibinfo{journal}{Phys. Rev. E} \textbf{\bibinfo{volume}{75}},
  \bibinfo{pages}{026102} (\bibinfo{year}{2007}).

\bibitem{stanley71} H. E. Stanley, Introduction to Phase Transitions and Critical
Phenomena (Oxford University Press, New York, 1971).

\bibitem[{\citenamefont{Feller}(1971)}]{feller71}
\bibinfo{author}{\bibfnamefont{W.}~\bibnamefont{Feller}},
  \emph{\bibinfo{title}{An introduction to probability theory and its
  applications}}, vol.~\bibinfo{volume}{II} (\bibinfo{publisher}{Wiley, New
  York}, \bibinfo{year}{1971}).


\bibitem[{\citenamefont{Cobham}(1954)}]{cobham54}
\bibinfo{author}{\bibfnamefont{A.}~\bibnamefont{Cobham}}, \bibinfo{journal}{J.
  Op. Res. Soc.} \textbf{\bibinfo{volume}{2}}, \bibinfo{pages}{70}
  (\bibinfo{year}{1954}).

\bibitem{hidalgo06} C. A. Hidalgo, Physica A {\bf 369}, 877-883 (2006).

\bibitem{vazquezPA07} A. Vazquez, Physica A {\bf 373}, 747 (2007).

\bibitem{goncalves08} B. Gon\c{c}alves and J. J. Ramasco, E-print: arxiv.org/abs/0803.4018.

\end{thebibliography}

\end{document}